\documentclass[acus]{JAC2000}   % US letter paper size
\usepackage{graphicx}     % CERN usage

\newcommand{\be}{\begin{equation}}
\newcommand{\ee}{\end{equation}}
\newcommand{\bea}{\begin{eqnarray}}
\newcommand{\eea}{\end{eqnarray}}

\newcommand{\bc}{\begin{center}}
\newcommand{\ec}{\end{center}}

\begin{document}
\title{QUADRUPOLE MISALIGNMENTS AND STEERING IN LONG LINACS
\thanks{Work supported by
DOE contract DE-AC03-76SF00515.}}
\author{G. V. Stupakov}
\maketitle

\begin{abstract}
We present a study of orbit jitter and emittance growth in a long
linac caused by misalignment of quadrupoles. First, assuming a FODO
lattice, we derive analytical formulae for the RMS deviation of the
orbit and the emittance growth caused by random uncorrelated
misalignments of all quadrupoles. We then consider an alignment
algorithm based on minimization of BPM readings with a given BPM
resolution and finite mover steps.
\end{abstract}

\section{Introduction}

In this paper we study the emittance dilution of a beam caused by
quadrupole misalignments in a long linac. To suppress the beam
break-up instability an energy spread is usually introduced in the
beam. For the Next Linear Collider (NLC) \cite{nlc_zdr96}, the rms
energy spread within the bunch will be of order of 1\%. Due to the
lattice chromaticity, the deflection of the beam by displaced
quadrupoles results in the dilution of the phase space and the growth
of the projected emittance.

The effect of lattice misalignments has been previously studied in
many papers. A qualitative analysis and main scalings were obtained
in Ref.\@ \cite{ruth87}, and detailed studies with intensive computer
simulations are described in Refs.\@
\cite{raubenheimer91r,sery96n,sery97m}. The purpose of this paper is
to develop a simple model based on a FODO lattice approximation for
the linac which allows an analytic calculation of the emittance
dilution. The model can be also generalized, to include a slow
variation of the lattice parameters, as well as variation of both
beam energy and the energy spread \cite{to_be_published}.

Throughout this paper we  assume that the number of quadrupoles in
the linac is large, $N \gg 1$, and neglect terms of the relative
order of $N^{-1}$ in the calculations. For future linear colliders
with the center of mass energy in the range of 1 TeV, typically
$N\sim 10^3$, and $N^{-1}$ is indeed a small number.

\section{Beam Orbit in Misaligned Lattice}
        \begin{figure}[ht]
        \begin{center}
        \includegraphics[scale=0.4]{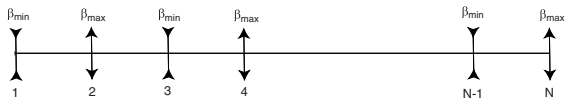}
        \end{center}
        \caption{FODO lattice of a linac. Beam positions are
        measured at the center of each quadrupole.
        \label{FODO}}
        \end{figure}
Let us consider a FODO lattice with a cell length $l$ and a phase
advance $\mu$ per cell, consisting of $N$ thin quadrupoles as shown
in Fig. \ref{FODO}. The focal length of the quadrupoles is equal to
$\pm F$ where the positive and negative values of $F$ refer to the
focusing and defocusing quadrupoles respectively. The beam is
injected in the linac at the center of the first quadrupole, at
$s=0$, with the zero offset and angle, and the beam emittance is
measured at the center of the last, $N$th, quadrupole. For the beam
position (horizontal or vertical) at the locations of the quadrupoles
we will use the notation  $x_1, x_2, \ldots, x_{N-1}, x_N$, and the
orbit angle at the center of the $k$th quadrupole is denoted by
$x'_k$. The initial conditions for the orbit are $x_1 = x'_1 =0$.
Note that due to our choice of positions, the derivative of the beta
function, and hence the Twiss parameter $\alpha$, at all locations 1
through $N$, are equal to zero.

We now assume that each quadrupole in the lattice is misaligned in
the transverse direction relative to the axis of the linac by
$\xi_i$, ($1 \le i \le N$), where $\xi_i$ are \emph{random,
uncorrelated} numbers. Due to the deflection by misaligned
quadrupoles, the original straight orbit will be perturbed. The
offset $x_i$ can be found as
    \be
    x_i = \sum_{k<i}R_{ik}\theta_k,
    \ee
where $R_{ik}$ is the $(1,2)$ element of the transfer matrix $R$ and
$\theta_i$ is the deflection angle resulting from the  offset of the
$i$th quadrupole, $\theta_i=\pm \xi_i/F$, for the focusing and
defocusing quadrupoles. We have $R_{ik} = \sqrt{\beta_i
\beta_k}\sin\Delta \psi_{ik}$, where the betatron phase advance
$\Delta \psi_{ik}$ between $k$th and $i$th quadrupoles ($k<i$) is
$\Delta \psi_{ik} = (1/2)(i-k)\mu$.

We will also need the orbit angles $x_i'$ where the
prime denotes the derivative with respect to the longitudinal
coordinate $s$. For $x_i'$ we have
    \be
    x_i' = \sum_{k \le i}G_{ik}\theta_k,
    \ee
where  $G_{ik}$, is the $(2,2)$ element of the transfer
matrix, $G_{ik} = \sqrt{\beta_k/\beta_i}\cos\Delta \psi_{ik}$,
(note that, due to our choice, $\alpha_i = 0$).

\section{RMS value for the beam offset}

To characterize the deviation of the orbit from the linac axis, we
will calculate the average value $\langle x_N^2\rangle$, where
the angular brackets denote averaging over all possible values of
$\xi$. We assume that the average offset $\langle \xi_i\rangle$
vanishes, hence $\langle x_N\rangle =0$.

For the lattice shown in Fig. \ref{FODO} the deflection angle
$\theta_k$ due to the misaligned $k$th quadrupole is given by $
\theta_k = (-1)^{k} \xi_k/F$, and the beam offset at the end of the
linac is
    \be \label{x_n}
    x_{N} = \sum_{k=1}^{N-1}R_{Nk}(-1)^{k}
    \frac{\xi_k}{F}.
    \ee
For the variance of $x_{N}$ we have
    \be \label{x_n_squared1}
    \langle x_{N}^2\rangle = \frac{\langle \xi^2\rangle}{F^2}
    \sum_{l=1}^{N-1}R_{Nl}^{2},
    \ee
where we have used $\langle \xi_k \xi_l\rangle =\langle \xi^2\rangle
\delta_{kl}$, with $\langle \xi^2\rangle$ being the variance of the
random variables $\xi_i$. To calculate the sum in Eq.
(\ref{x_n_squared1}), one can average $R_{Nl}^{2}$ over the betatron
phase value $R_{Nl}^{2}\rightarrow \frac{1}{4}\beta_N \beta_l$. One
finds,
    \be \label{rms_x_n}
    \frac{\langle x_{N}^2\rangle}{\beta_{N}}
    = \frac{N \langle \xi^2\rangle}{4 F^2}
    (\beta_{\max}+\beta_{\min})
    =4N\frac{\langle \xi^2\rangle}{ l}
    \tan \frac{\mu}{2}.
    \ee
We see that the rms value $\langle x_{N}^2\rangle^{1/2}$ scales as
$N^{1/2}$, which is a characteristic feature of the random walk
motion.

In a similar fashion, one can find the rms angular spread orbits
$\langle {x'}_N^2\rangle^{1/2}$ at the end of the linac. Starting
from the general expression
    \be \label{x_n_prime}
    x'_N = \sum_{k=1}^{N}G_{Nk}(-1)^{k-1}
    \frac{\xi_k}{F},
    \ee
and performing the same averaging as for derivation of Eq.
(\ref{rms_x_n}), one finds $\langle {x'}_N^2\rangle =\langle
x_N^2\rangle/\beta_N^2$, where $\langle {x}_N^2\rangle$ is given by
Eq. (\ref{rms_x_n}).

\section{Chromatic Emittance Growth}\label{sec2}

When the beam has a nonzero energy spread, due to the chromaticity of
the lattice, the misalignments cause an effective emittance growth of
the beam \cite{ruth87}. We will calculate the emittance increase,
assuming that the beam energy $E$ and the relative energy spread in
the beam $\delta$  are constant. We will also assume that the
resulting emittance growth is much smaller than the initial emittance
of the beam. In this case, we can use the following formula for the
final emittance growth
    \bea \label{emit_increase}
    \Delta \epsilon
    &=& \frac{1}{2}
    \left[
    \beta_{N}^{-1}
    \langle(\Delta x-\langle \Delta x \rangle)^2\rangle_{\xi\delta}
    \right.
    \nonumber \\
    &+& \left.\beta_{N}
    \langle(\Delta x'-\langle \Delta x' \rangle)^2\rangle_{\xi\delta}
    \right],
    \eea
where $\Delta x$ and $\Delta x'$ are the spread in the coordinate and
the angle within the bunch at the and of the linac, and the angular
brackets with the subscript ${\xi\delta}$ denote a double averaging:
first, averaging over the random misalignment of the quadrupoles and
then averaging over the energy distribution function in the beam. We
will assume that the energy spread in the beam $\delta$ is so small,
that one can use a linear approximation for calculation of $\Delta x$
and $\Delta x'$, $\Delta x = \delta \cdot x_{N\delta} \equiv \delta
\cdot\partial x_{N}/\partial \delta$ and  $\Delta x' = \delta \cdot
x'_{N\delta} \equiv \delta \cdot\partial x'_{N}/\partial \delta$.
Since $\langle \xi_i \rangle = 0$, hence $\langle \Delta x \rangle =
\langle \Delta x' \rangle = 0$.  In this approximation Eq.
(\ref{emit_increase}) reduces to
    \be \label{eq16}
    \Delta \epsilon
    = \frac{1}{2} \overline{\delta^2}
    \left(
    \beta_{N}^{-1}
    \left\langle
    x_{N\delta}^2
    \right\rangle
    + \beta_{N}
    \left\langle
    {x'}_{N\delta}^2
    \right\rangle
    \right),
    \ee
where $\overline{\delta^2}$ is the variance of the energy spread
within the beam.

To calculate $x_{N\delta}$ and $x'_{N\delta}$ we need to take the
derivatives of  Eqs. (\ref{x_n}) and (\ref{x_n_prime}) with respect
to $\delta$. For a long linac, the dominant contribution to $\Delta
\epsilon$ comes from the dependence of the phase advance $\Delta
\psi_{ik}$ versus energy, so we need to differentiate only
$\sin\Delta \psi_{ik}$ (or $\cos\Delta \psi_{ik}$) terms in the sum.
Calculation gives
    \be \label{dispers}
    \beta_{N}^{-1}
    \left\langle
    x_{N\delta}^2
    \right\rangle
    =
    \beta_{N}
    \left\langle
    {x'}_{N\delta}^2
    \right\rangle
    =
    \frac{4}{3}
    N^3
    \frac{\langle \xi^2 \rangle}{l}
    \tan^3\frac{\mu}{2}
    \,.
    \ee
which gives for the emittance dilution
    \be \label{emit_growth}
    \Delta \epsilon
    =
    \frac{4}{3}\overline{\delta^2}
    N^3\frac{\langle \xi^2 \rangle}{l}
    \tan^3\frac{\mu}{2}
    \,.
    \ee
As we see, the increase in the emittance scales with the number of
quadrupoles as $N^3$.

In the above derivation, to find the dispersion of the beam at the
end of the linac, we explicitly differentiated Eq.~(\ref{x_n}) with
respect to the energy. One can use another formula for computing
$\partial x_{N}/\partial \delta$ \cite{to_be_published},
    \be \label{another_exp_for_dispersion}
    \frac{\partial x_{N}}{\partial \delta}
    = \sum_{k=1}^{N-1}R_{Nk}(-1)^{k}
    \frac{x_k-\xi_k}{F},
    \ee
that takes into account that the dispersion is generated due to the
offset of the particle relative to the center of the quadrupole, and
propagates downstream with the same matrix element $R_{Nk}$.

\section{Very long linac}\label{sec5}

Increasing the length of the linac and the number of quadrupoles $N$
brings us to the regime where Eq. (\ref{emit_growth}) is not valid any
more.  The transition occurs when the phase advance over the
length  of the linac due to the energy variation $\delta$ becomes
comparable to $\pi/2$, $N \delta \cdot d \mu/d\delta \sim \pi/2$. In
this case, the differential approximation $\Delta x = \delta \cdot
\partial x_{N}/\partial \delta$ that was used in Section \ref{sec2}
is not valid any more, and the scaling $\Delta \epsilon\propto N^3$
breaks down.

We can estimate the emittance dilution in this regime, using the
following arguments. Let us denote by $l_c$ the decoherence length in
the linac such that $(l_c/l)  \delta \cdot d\mu/d\delta \sim \pi/2$ ($l$
is the FODO cell length). When the beam passes the distance $l_c$,
due to filamentation, the betatron oscillations of the beam are
converted into the increased emittance, and the subsequent motion
becomes uncorrelated with the previously excited betatron
oscillations. The emittance growth on the distance $l_c$ is given by
Eq. (\ref{emit_growth}), in which $N=2l_c/l$,
    \be
    \Delta \epsilon_c
    =
    \frac{4}{3}\overline{\delta^2}
    \left(\frac{2l_c}{l}\right)^3
    \frac{\langle \xi^2 \rangle}{l}
    \tan^3\frac{\mu}{2}
    \approx
    \frac{\langle \xi^2 \rangle}{l\sqrt{\overline{\delta^2}}}
    \,.
    \ee
The total emittance increase in the  linac  of length $l_L$ in this
regime is equal to $\Delta \epsilon_c$ multiplied by the number of
coherent distances $l_L/l_c$ in the linac
    \be
    \Delta \epsilon
    =
    \Delta \epsilon_c \frac{l_L}{l_c}
    \sim
    \frac{l_L \langle \xi^2 \rangle}{l^2\tan\frac{\mu}{2}}
    \sim
    \frac{N \langle \xi^2 \rangle}{l\tan\frac{\mu}{2}}
    \,.
    \ee

Note that if the linac length $l_L<l_c$, the emittance dilution is
reversible in principle -- the initial beam emittance can be
recovered by taking out the dispersion generated by the misaligned
quadrupoles downstream of the linac. For very long linacs, when
$l_L>l_c$, the emittance growth becomes irreversible due to the phase
space filamentation.

\section{Alignment with account of BPM errors and finite mover steps}

Measuring the beam position at each quadrupole, with the knowledge of
the lattice functions, allows us to find the quadrupole offsets
$\xi_i$. Moving the quadrupoles by distance $-\xi_i$ would position
them in the original state, and restore the ideal lattice. Of course,
in reality, there are many factors, such as wakefields and
measurement errors, that do not allow to perfectly align the lattice.
Here we will study two such effects: errors associated with the BPM
measurements, and finite step of the quadrupole movers.

Consider first the effect of BPM errors. Due to the finite resolution
of BPMs the measured vector of the beam transverse offsets
$X^M=(x_{1}^M \ldots x_{N}^M)$ differs from the exact values
$X=(x_{1} \ldots x_{N})$ by an error vector $e$, $X^M=X+e$, where $e
=(e_{1} \ldots e_{N})$. The errors are small relative to the measured
values, $|e_i| \ll |x_i|$. We assume that the BPMs are built in the
quadrupoles, and the quadrupole displacement $\xi_k$ also moves the
center line of the BMP, so that BPM reading is $x_k^M=x_k-\xi_k+e_k$.
Using the measured offsets $x_k^M$ we infer the quadrupole offsets
$\zeta_k$ from the following equation
    \be\label{eq_for_zeta}
    x_i^M+\zeta_i
    =\sum_{k=1}^{i-1} R_{ik}(-1)^k\frac{\zeta_k}{F}.
    \ee
Note that without errors, $e_k=0$, we would find from Eq.
(\ref{eq_for_zeta}) the correct value $\zeta_k=\xi_k$. Measurement
errors $e_k$ cause the inferred values of the offsets differ from the
true ones, $\zeta_k\ne\xi_k$.

We then align the lattice by moving the quadrupoles by distance
$-\zeta_k$. After the alignment the corrected beam orbit $\tilde x_i$
does not vanish:
    \bea
    \tilde x_i
    &=&\sum_{k=1}^{i-1} R_{ik}(-1)^k\frac{\xi_k-\zeta_k}{F}
    \nonumber \\
    &=&x_i-x^M_i-\zeta_i
    =-e_i+\xi_i-\zeta_i.
    \eea
Since the quadrupoles after alignment are located at $\xi_k-\zeta_k$,
the beam offset \emph{relative to the center of the quadrupole},
$\tilde x_k-(\xi_k-\zeta_k)$, is equal to $-e_k$. This allows us to
use Eq. (\ref{another_exp_for_dispersion}) to find the emittance
dilution in the linac after the alignment,
    \be
    x_{N\delta}
    = -\sum_{k=0}^{N-1}R_{Nk}(-1)^{k-1}
    \frac{e_k}{F}.
    \ee
Assuming that  $e_k$ are uncorrelated random numbers makes the
problem equivalent to the orbit equation (\ref{x_n}) with the result
given by Eq. (\ref{rms_x_n}),
    \be \label{rms_eta_n}
    \beta_{N}^{-1}
    \left\langle
    x_{N\delta}^2
    \right\rangle
    =4N\frac{\langle e^2\rangle}{l}
    \tan \frac{\mu}{2}.
    \ee
We see that the rms value of the dispersion at the and of the linac
after alignment scales as $\sqrt{N}$. Calculating in a similar way
the variance of the derivative $x'_{N\delta}$, gives the chromatic
emittance growth after alignment,
    \be \label{emit_growth_after_alignment}
    \Delta \epsilon
    =4N\overline{\delta^2}\frac{\langle e^2\rangle}{l}
    \tan \frac{\mu}{2}
    \,.
    \ee

Let us now assume that in addition to the BPM errors the quadrupole
movers have a finite step so that the final position of the
quadrupoles $\zeta_k$ after alignment is  $\xi_k-\zeta_k + r_k$,
where as above, $\zeta_k$ is the offset inferred from the
measurements (and containing BPM errors), and $r_k$ is the quadrupole
movement error. Again, we assume that $r_k$ are random, uncorrelated
numbers, and of course uncorrelated with the BPM errors $e_k$.
For the beam orbit after alignment we now have
    \be
    \tilde x_i
    =-e_i + \xi_i-\zeta_i+\sum_{k=1}^{i-1} R_{ik}r_k
    \ee
with the resulting emittance growth that is a combination of Eqs.
(\ref{emit_growth_after_alignment}) and (\ref{emit_growth}),
    \be \label{emit_growth_both_errors}
    \Delta \epsilon
    =4N\overline{\delta^2}\frac{\langle e^2\rangle}{l}
    \tan \frac{\mu}{2}
    +
    \frac{4}{3}\overline{\delta^2}
    N^3\frac{\langle r^2 \rangle}{l}
    \tan^3\frac{\mu}{2}
    \,.
    \ee
From this equation, it follows that for a large $N$, the contribution
of the movers errors becomes more important and imposes tighter
tolerances on the movers.

\end{document}